\documentclass[aps, twocolumn, nofootinbib, showpacs, prl]{revtex4}
\usepackage{graphicx}
\usepackage{amsfonts}
\usepackage{amssymb}
\usepackage{amsbsy}
\usepackage{amsmath}
\usepackage{latexsym}
\usepackage{natbib}
\usepackage{bm}
\usepackage{color}
\usepackage{float}

\begin{document}
\title{The River Model of Gravitational Collapse}

\author{Soumya Chakrabarti}
\email{soumya.chakrabarti@vit.ac.in}

\affiliation{School of Advanced Sciences, Vellore Institute of Technology, Tiruvalam Rd, Katpadi, Vellore, Tamil Nadu 632014, India}


\date{\today}

\begin{abstract}
We show that the transformation of a time-evolving spherically symmetric metric tensor into a Painlev\'e-Gullstrand-Lema\^itre form brings forth a few curious consequences. The time evolution describes a non-singular gravitational collapse, leading to a bounce and dispersal of all the clustered matter, or a wormhole geometry for certain initial conditions. The null convergence condition is violated only at the onset of bounce or the wormhole formation. As an example, the requirements to develop a Simpson-Visser wormhole/regular black-hole geometry is discussed. The solution can be regarded as a new time-evolving twin of \textit{sonic dumb holes} found in analog gravity.
\end{abstract}

\maketitle

The modern idea of gravitational physics is based on an intuitive interpretation of the laws of nature and a few paradoxes. General Theory of Relativity (GR) provides a way to address them based on a geometric description. Some of these paradoxes have developed popular research problems over the years and they can be classified depending on their origin and basic motivations. There are problems which do not necessarily require a gravitational environment, for example, the study of topological defects evolving from the residues of cosmological phase transitions \cite{topo}. They can carry signatures of an early cosmic expansion history. Focus must equally be given on some problems strongly related to the background gravitational environment, for example, the dynamics of a collapsing stellar matter distribution after the death of a star. It is widely believed that such a gravitational collapse will produce a zero proper volume singularity which will \textit{probably} remain hidden behind a null surface, known as the \textit{horizon} \cite{os}. A formation of horizon indicates a black hole from which information cannot escape, atleast classically. It is a natural intuition to imagine that near a \textit{zero proper volume} quantum effects will generate some modifications and lead to phenomena like Hawking radiation \cite{hawking}. However, no such complete model of gravitational collapse has been proposed till date. The formation of a horizon itself remains a debatable issue \cite{penrose, censor} and has led to a number of proposals and counter-proposals, the most remarkable amongst all being the \textit{`cosmic censorship conjecture'}. Once again, a quantum correction can perhaps provide a better understanding of how a horizon develops, however, any such correction is based on a quantum field theory that has found success only on very small scales. No successful analogue system with horizons have been constructed in the lab so far that can test quantum corrections on an appropriate scale or provide any alternative notion. In a sense this lack of explanation keeps an open end and demands new perspectives.

\begin{figure}
{\centerline{{\scalebox{0.37}{\includegraphics{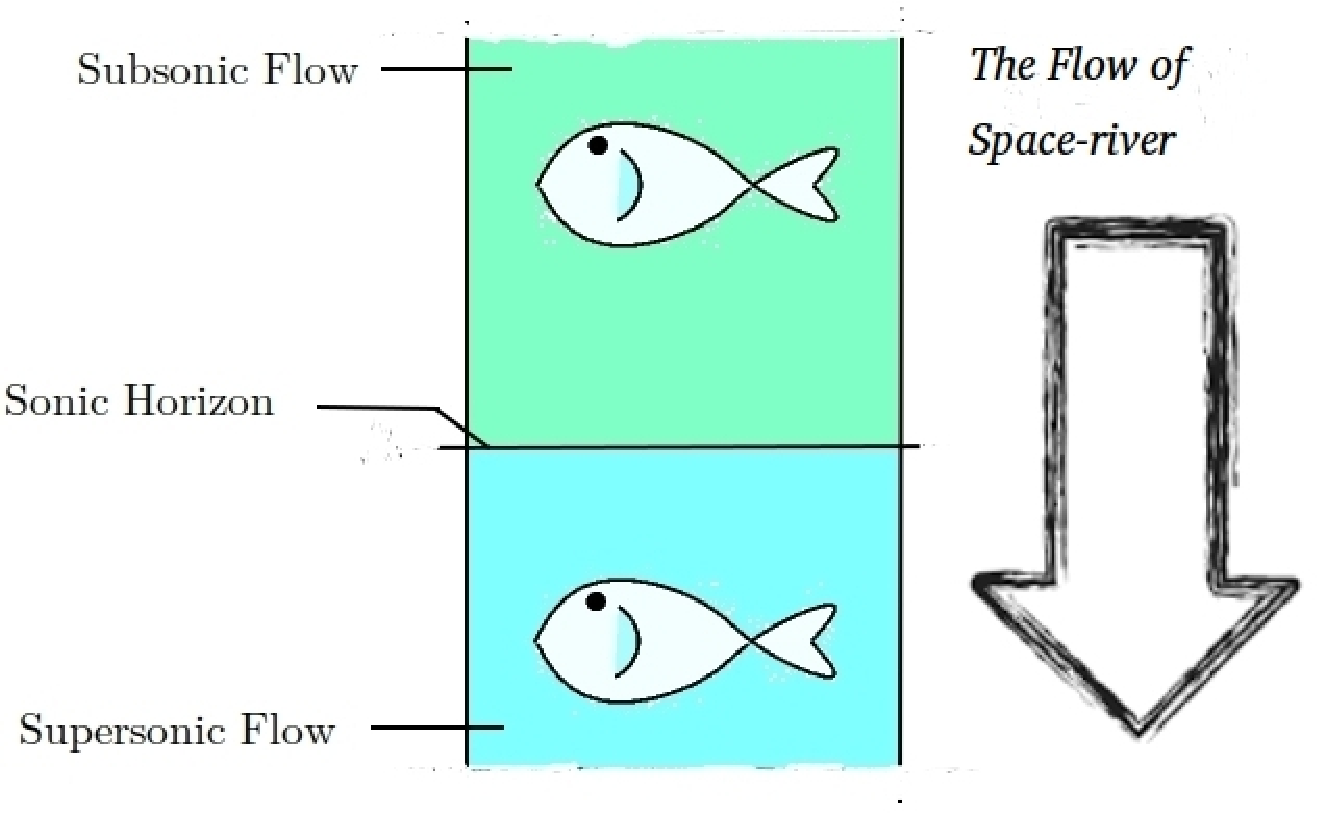}}}}}
\caption{\small \label{fig1} A simple visualization of a sonic black (dumb) hole. The water flow replicates space flowing towards a black hole. The fish in a supersonic region neve receives any sound from a second fish in the subsonic region as any emitted sound travels too slowly to propagate upstream. A sonic horizon forms at the location where the fluid velocity becomes supersonic.}
\end{figure}

It might be beneficial for relativists if the mathematics of classical gravitational collapse do not necessarily generate singularities for generic spacetime geometries. However, the known few models of non-singular collapse either relies on exotic matter distributions \cite{branden}, quantum corrections \cite{hayward, bojo} or suffers due to lack of physical motivation. In this letter we try to provide an escape clause, based on the concept of an analog black hole. We propose that the collapse of a stellar distribution can lead to non-singular outcomes if it has a time evolving analogue black hole structure. An analogue black hole, first proposed by Unruh \cite{unruh}, is written using a static spherically symmetric solution of Einstein field equations in Painlev\'e-Gullstrand-Lema\^itre (PGL) form  \cite{painleve, gulstrand}. We construct a time evolving PGL geometry to describe a gravitational collapse of sufficiently massive stellar distributions. There is no curvature singularity, no formation of horizon and therefore, the dichotomy related to cosmic censorship is avoided. The geometry is better explained by a sonic flow separated into downstream region and upstream region. These two regions are characterized by supersonic and subsonic speeds of the flow, respectively (shown in Fig. \ref{fig1}). A fish flowing along the supersonic downstream can never send a sound signal to a second fish flowing along the upstream. In effect, the boundary of the two regions work like a \textit{sonic event horizon}. Since the upstream region does not receive any sound across the horizon, this region acts like a \textit{`sonic dumb hole'}, an analogue black hole.

If a river consists of irrotational fluid (density $\rho$, pressure $p = p(\rho)$, velocity $v$) of negligible viscosity, it should obey the Navier-Stokes equations 
\begin{eqnarray}
&& {\nabla} \times v = 0 ~,~ \frac{\partial \rho}{\partial t} + {\nabla}\cdot (\rho {v}) = 0, \\&&
\frac{\partial {v}}{\partial t} + {v}\cdot {\nabla} {v} = -\frac{1}{\rho}{\nabla}p - {\nabla}\Phi.
\end{eqnarray}

In a general relativistic scenario $\Phi$ is equivalent to the gravitational potential. For fluctuations of the background flow as in $\rho = \rho_0 + \delta \rho$ and $v = v_0 + {\nabla}\phi$, the velocity fluctuation $\phi$ obeys \cite{unruh}
\begin{equation}
\nabla_\mu \nabla^\mu \phi = \frac{1}{\sqrt{-g}} \partial_\mu ( \sqrt{-g} g^{\mu \nu} \partial_\nu \phi ) = 0.
\label{waveeq}
\end{equation}

The background fluctuations correspond to a motion written using the metric
\begin{equation}
ds^2 = (c_s^2 - v_0^2) dt^2 + 2 v_0 dt dr - dr^2 - r^2 d\Omega^2. 
\label{metric}
\end{equation}

This is the standard PGL metric describing a stationary, spherically symmetric black hole. A horizon develops when the fluid velocity is equal to the sound speed, i.e., $v_0 = c_s$. The analogy becomes fascinating once the Schwarzschild metric is written in this form
\begin{equation}
\label{gullstrandmetric}
ds^2 = d{t_{eff}}^2 - (dr + v_0 d{t_{eff}})^2 - r^2 d\Omega^2.
\end{equation}
$v_0$ turns out to be the Newtonian escape velocity for a spherical object of mass $M$, $v_0 = \left(\frac{2GM}{r} \right)^{1/2}$. An object that starts to fall radially inward, freely from infinity towards the black-hole, records time as $t_{eff}$. The usual Schwarzschild time $t$ is related to this time by
\begin{equation}
t_{eff} = t + 2 (r r_s)^{1/2} + r_s \ln \left| {r^{1/2} - r_s^{1/2} \over r^{1/2} + r_s^{1/2}} \right|.
\end{equation}

The PGL analog of stationary black hole can be called a `River Model' (the name first coined in \cite{hamilton}). It comes from an imagination that space is flowing with Newtonian escape velocity, radially inwards, through a flat background. An object defined in this metric is moving alongwith this flow, obeying the laws of special relativity. There is an event horizon whenever the infall velocity is equal to the speed of light. Beyond this horizon all objects are carried away towards the central singularity with an infall velocity greater than the speed of light. The illustration in Fig. \ref{fig1} is done by comparing a couple of fishes in this current with photons. The upstream region is the exterior region where the \textit{`photon-fishes'} can still move against the flow. However, in the downstream region/interior, the inward flow is too fast (greater than the speed of light!) for any fish/photon to make a cross-over into the upstream and inevitably, they should fall towards a central singularity. Compared to the standard Newtonian picture contemplated in the works of Michell \cite{michell}, the river model is a non-conservative narrative. The geometry of a PGL metric has generated some limited curiosity from time to time within the community of relativists \cite{review}. However, Unruh's construction \cite{unruh1} is undoubtedly the most important of all as it inspired the foundation of \textit{`analog gravity'} \cite{barcelo, others}. In an analog gravity framework one works with a fluid flowing with pre-assigned velocity and simulates dynamical evolutions in a general relativistic spacetimes. It remains one of the few ways to explore gravity experimentally near quantum scales, using sonic analogs.

It has been discussed before that to admit a river analog (or a PGL form) a stationary black hole metric must be spatially flat at any fixed time, up to a conformal factor \cite{garat, doran}. This conformal factor includes the velocity of the river flowing through the background flat space. However, these claims apply only for a static or stationary metric. It is not possible to construct a model of stellar collapse using this metric unless one can find a time-evolving analogue of the PGL metric, which, till date has never been prescribed. We construct such a metric and argue that it describes a \textit{`river model'} of gravitational collapse. For a generic spherically symmetric metric we derive the transformation equations which the metric components should obey in order to have a PGL form. We find an exact solution, albeit a special case but nonsingular, which allows the collapse to either go on for an infinite time or forces a bounce and dispersal. We further discuss that for some reasonable initial radial profiles of the collapsing distribution the time-evolving metric components satisfy a Wormhole throat condition. In other words, a river-frame gravitational collapse can evolve into a Wormhole geometry.

We can write a general spacetime metric to describe a spherical star as $ds^{2}_{-} = A^2 dt^2 - B^2 dr_{c}^2 + r_{c}^2 C^2 d\Omega^2$. The interior of the spherical star contains a fluid with local anisotropy and heat flux, therefore, $T_{\alpha\beta}=(\rho+p_{t})u_{\alpha}u_{\beta}-p_{t}g_{\alpha\beta}+ (p_r-p_{t})\chi_{\alpha} \chi_{\beta}+q_{\alpha}u_{\beta}+q_{\beta}u_{\alpha}$. $\rho$, $p_{t}$ and $p_r$ are density, tangential and radial pressure and $q^{\alpha}=(0,q,0,0)$ is the radial heat flux. The four-velocity and the unit four-vector in radial direction follow usual normalizations. We introduce the transformation $r_{c}C = r$, such that
\begin{equation}\label{transformation}
dr_{c} = \frac{1}{C}(dr-r_{c}dC) = \frac{1}{C}\left\lbrace dr - r_{c}(\dot{C}dt + C'dr)\right\rbrace. 
\end{equation}
A dot represents derivative with respect to $t$ and a prime is derivative with respect to $r$. The transformation allows us to write the spherical metric in the following form

\begin{eqnarray}\label{interior}\nonumber
&&ds^{2}_{-} = \Bigg(A^{2} - \frac{r^{2}B^{2}\dot{C}^{2}}{C^{4}}\Bigg)dt^{2} + \Bigg(\frac{2rB^{2}\dot{C}}{C^{3}} - \frac{2r^{2}B^{2}\dot{C}C'}{C^{4}}\Bigg)\\&&
dtdr - \Bigg(\frac{B^2}{C^2} - \frac{2rB^{2}C'}{C^3} + \frac{r^{2}B^{2}C'^{2}}{C^{4}}\Bigg)dr^{2} - r^{2}d\Omega^{2}. 
\end{eqnarray}

Comparing the above metric with a generic PGL form
\begin{eqnarray}\label{gen_PG}
ds^2 = (1-\zeta^2)dt^2 \pm 2\zeta drdt - dr^2 - r^2 d\Omega^{2},
\end{eqnarray}

the original metric components and $\zeta(r,t)$ should satisfy the following set of differential equations,
\begin{eqnarray}\label{formationeq}
&& \pm 2\zeta = \frac{2rB^{2}\dot{C}}{C^{3}} - \frac{2r^{2}B^{2}\dot{C}C'}{C^{4}}, \\&&
1-\zeta^2 = A^{2} - \frac{r^{2}B^{2}\dot{C}^{2}}{C^{4}}, \\&&
\frac{B^2}{C^2} - \frac{2rB^{2}C'}{C^3} + \frac{r^{2}B^{2}C'^{2}}{C^{4}} = 1.
\end{eqnarray}

We find a particular exact solution of these equations as follows

\begin{eqnarray}\nonumber
&& A(r,t)^{2} = \Bigg[1 + \alpha(r)^{2}\beta^{2}\Bigg\lbrace \Bigg(1 \pm \frac{4r^{5/2}}{\alpha(r)^{1/2}}\Bigg) - \Bigg(1 \pm \\&&\nonumber
\frac{2r^{5/2}}{\alpha(r)^{1/2}}\Bigg)^{2} \Bigg(1-\frac{r}{(\alpha(r)^{1/2} + 2r^{1/2})^{2}}\Bigg) \Bigg\rbrace\Bigg]\Big[\delta - g T(t)^n \Big],\\&&\label{B0}
B(r,t)^{2} = r \alpha(r) T(t)^{2} e^{\pm 4 f(r)}, \\&&\label{B}
C(r,t)^{2} = r^{2} T(t)^{2}e^{\pm 4 f(r)}, ~~~ f(r) = \int\frac{r^{\frac{3}{2}}}{\alpha(r)^{\frac{1}{2}}}dr, \\&&\label{C}
\zeta(r,t) = \mp 2\alpha(r)^{1/2}\beta r^{5/2}(\delta - gT^{n})^{1/2}, \\&&\label{zetaform}
T(t) = \Bigg(\frac{\delta}{g}\Bigg)^{1/n}\Bigg[1 - tanh \Bigg\lbrace \frac{\delta n^{2}}{4}(\beta t - C_{1})^{2}\Bigg\rbrace\Bigg]^{1/n}. \label{Tform}
\end{eqnarray} 

$\beta$, $\delta$ and $g$ are positive parameters. $\alpha(r)$ is a positive but otherwise arbitrary function of $r$. For a constant $\alpha(r) = \alpha_0$, the function $f(r)$ is simplified and the the metric coefficients $B$ and $C$ are written as
\begin{eqnarray}
&& B(r,t)^{2} = \alpha_0 r T(t)^{2} e^{\pm \frac{8}{5}{\alpha_0}^{-1/2}r^{5/2}}, \\&&
C(r,t)^{2} = r^{2} T(t)^{2} e^{\pm \frac{8}{5}{\alpha_0}^{-1/2}r^{5/2}}.
\end{eqnarray}

It is evident form Eq. (\ref{zetaform}) that for a real metric function $\zeta$, $(\delta - gT^{n}) > 0$. Moreover, Eq. (\ref{Tform}) indicates that the radius of the two-sphere $C(r,t)^{2}$ can only be zero when the hyperbolic tangent function tends to $1$. Solutions for both $n > 0$ and $n < 0$ are allowed, however, they are of different physical nature. A $n > 0$ case shows a forever collapsing spherical star reaching a zero proper volume only at $t \rightarrow \infty$. On the other hand, a $n < 0$ case is a collapse and bounce case, without any formation of zero proper volume singularity. \\

We assume the exterior region of the collapsing star to be a Schwarzschild metric
\begin{equation}\nonumber
ds^2 = (1-\zeta^2)dt_s^2 - \frac{dr^2}{1-\zeta^2} - r^2 d\Omega^2 ~,~ \zeta = \pm \sqrt{\frac{R}{r}}, ~ R = 2m.
\end{equation}
Taking $t_s = t + g(r)$ the metric can be transformed into
\begin{equation}
ds^2 = (1-\zeta^2)dt^2 \pm 2\zeta drdt - dr^2 - r^2 d\Omega^2, \label{PG}
\end{equation}
which is a PGL-compatible form, provided
\begin{equation}
g^{\prime} = \pm \frac{\zeta}{1-\zeta^2} ~,~ g = \mp R \left( 2\sqrt{\frac{r}{R}} + \ln{\frac{\sqrt{r}-\sqrt{R}}{\sqrt{r}-\sqrt{R}}} \right).
\label{PGtSolution}
\end{equation}
Now, both the interior and the exterior of the collapsing sphere are in a single metric form Eq. (\ref{gen_PG}) with a single metric function 
\begin{eqnarray}\label{GeneralFluidpsi}
\zeta = \left\{ \begin{array}{ll}
\mp 2\alpha^{1/2}\beta r^{5/2}(\delta - gT^{n})^{1/2}, \\
- \sqrt{R/r}.
\end{array} \right. 
\end{eqnarray}

We refer to this metric and coordinate transformations as a generalized PGL form. The interior of such a collapsing star can describe some interesting geometric features. To explore this we construct an embedding geometry for a generic metric of pattern $ds^{2} = T(t)^{2} [A(r)^{2}dt^{2} - B(r)^{2}dr^{2} - r^{2}d\Omega^2]$, of which the generalized PGL metric is a special case. On a the spatial slice of constant $t$ and $\theta = \pi/2$ the metric looks like
\begin{equation}\label{3_d}
dl^2 = B^{2}T^{2}dr^{2} + r^{2}T^{2}d\phi^{2}.
\end{equation}
$dl^2$ is the metric on a surface of revolution $\rho = \rho(z)$ embedded in a three-dimensional space with an Euclidean metric $dl^2 = dz^2 + d\rho^2 + \rho^2 d\phi^2$, where $z$, $\rho$ and $\phi$ are cylindrical coordinates. Comparing with Eq. (\ref{3_d})
\begin{equation}
\rho^2 = r^{2}T^{2}, ~~ dz^{2} + d\rho^{2} = B^{2}T^{2}dr^{2}.
\end{equation}

For a constant $t$, $d\rho = T dr$ and
\begin{equation}\label{ee2}
\frac{d\rho}{dz} = \frac{1}{(B^{2}-1)^{\frac{1}{2}}}, ~~ \frac{d^2\rho}{dz^2} = -\frac{B'B}{T(B^{2}-1)^{2}}.
\end{equation}

\begin{figure}
{\centerline{{\scalebox{0.70}{\includegraphics{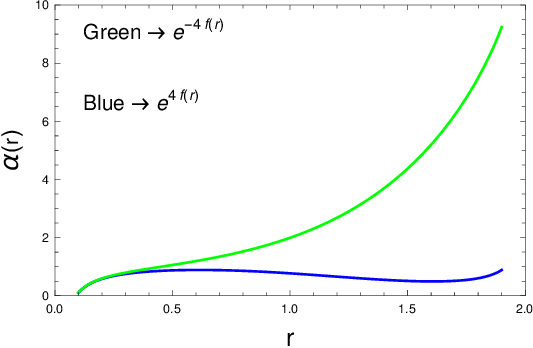}}}}}
{\centerline{{\scalebox{0.70}{\includegraphics{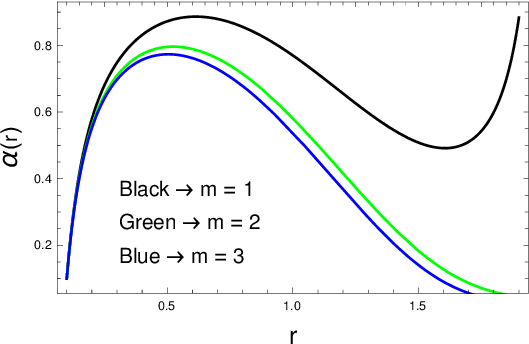}}}}}
\caption{\small \label{fig2} Numerical solutions of $\alpha(r)$ for different signs of $e^{\pm 4 f(r)}$ (top graph) and different values of $m$ (bottom graph).}
\end{figure}

This formulation leads us to check if a Wormhole throat condition is satisfied during the collapse or not. A throat is a two-dimensional space-like surface having the projected shape of a sphere, located at a certain minima $r = r_{w}$ \cite{throat}. On an embedding diagram, this sphere of $r = r_{w}$ is represented by a circle of radius $\rho$ on the surface of revolution. An usually tube-shaped wormhole can have a throat acting as pathway between two universes. Radial null rays converge and become parallel at a wormhole throat before eventually diverging on the other side. Naturally, at the throat the radius of the circle $\rho(z)$ has a minimum. Conditions for a formation of this minimum is
\begin{equation}\label{ee4}
\frac{d\rho}{dz}\Big|_{r_{w}} = 0, ~~ \frac{d^2\rho}{dz^2}\Big|_{r_{w}} > 0.
\end{equation}

Using Eqs. (\ref{B0}), (\ref{B}) and (\ref{C}), for a constant $\alpha = \alpha_0$, the first wormhole throat condition Eq. (\ref{ee4}) becomes
\begin{equation}
\frac{d\rho}{dz}\Big|_{r_{w}} = \frac{1}{\alpha_0 r T(t)^{2} e^{\pm \frac{8}{5}{\alpha_0}^{-1/2}r^{5/2}} - 1}\Big|_{r_{w}}.
\end{equation}
It can only be zero at $r_{w} \rightarrow \infty$, which means a wormhole is never developed in finite time. However, for the more generic case $\alpha = \alpha(r)$, 
\begin{equation}
\frac{d\rho}{dz}\Big|_{r_{w}} = \frac{1}{\alpha(r) T(t)^{2} e^{\pm 4 f(r)} - 1}\Big|_{r_{w}},
\end{equation}
and there is a real possibility that the first wormhole condition is satisfied for a real and finite value of $r$. It depends on the functional form of $k(r)$ or more precisely, $\alpha(r)$. We demonstrate this through an example. Recently it has been proved that a wide class of Wormhole solutions can develop out of a gravitational collapse of imperfect fluid \cite{scsk}. One such example is a Simpson-Visser metric, which, for different values of a parameter $a$ can represent different geometric structures (a Schwarzschild metric for $a = 0$, a traversable wormhole for $a > 2m$, one-way wormhole for $a = 2m$; see \cite{simpson, scsk} for more detailed discussions). In a coordinate range $r\in(r_w,+\infty)$, where $r_w$ is the wormhole throat the metric can be written as 
\begin{equation}\label{beginning}
ds^{2} = \Big(1-\frac{2m}{r}\Big)dt^{2} - \frac{dr^{2}}{\Big(1 - \frac{a^{2}}{r^{2}}\Big)\Big(1-\frac{2m}{r}\Big)} - r^{2} d\Omega^2.
\end{equation}

If we expect the river model of gravitational collapse to produce a Simpson-Visser wormhole, the metrics should be comparable on the spatial slice of constant $t$ and $\theta = \pi/2$, i.e., coefficients of $g_{11}$ should match, leading to
\begin{eqnarray}\label{wormholecond2}
\frac{\alpha'}{\alpha} + \frac{1}{r} \pm 2r^{\frac{3}{2}}\alpha^{-\frac{1}{2}} + \frac{\frac{2a^2}{r^3}}{(1-\frac{a^2}{r^2})} - \frac{\frac{2m}{r^2}}{(1-\frac{2m}{r})}.
\end{eqnarray}

If $a = 0$, the collapse should end up in a Schwarzschild black hole. In this limit, it is possible to solve Eq. (\ref{wormholecond2}) analytically and write 
\begin{eqnarray}\label{alphaform}\nonumber
&&\alpha(r)\Big|_{BH} = \frac{1}{36 (2m-r)^2} \Bigg[36 m^6 r-12 m^5 r^2-59 m^4 r^3 \\&&\nonumber
+34 m^3 r^4 + 21 m^2 r^5 - 20 m r^6 + 4 r^7 + 12 m^{11/2} r^{3/2} \\&&\nonumber
\sqrt{\frac{2 m-r}{m}} sin^{-1}\Big(\frac{\sqrt{r}}{\sqrt{2} \sqrt{m}}\Big) - 72 m^{13/2}\sqrt{r} \sqrt{\frac{2m-r}{m}} \\&&\nonumber
sin^{-1}\Big(\frac{\sqrt{r}}{\sqrt{2} \sqrt{m}}\Big) + 72 c_1 m^{7/2} \sqrt{\frac{2 m-r}{m}} \sqrt{2m-r} \\&&\nonumber
sin^{-1}\Big(\frac{\sqrt{r}}{\sqrt{2} \sqrt{m}}\Big) - 72 c_1 m^3 \sqrt{r} \sqrt{2m-r} + 12 c_1 m^2 r^{3/2} \\&&\nonumber
\sqrt{2m-r} + 60 c_1 m r^{5/2} \sqrt{2m-r} - 24 c_1 r^{7/2} \sqrt{2m-r} \\&&\nonumber
+ 72 {c_1}^2 m - 36 {c_1}^2 r + 60 m^{9/2} r^{5/2} \sqrt{\frac{2 m-r}{m}} \\&&\nonumber
sin^{-1}\Big(\frac{\sqrt{r}}{\sqrt{2} \sqrt{m}}\Big) -24 m^{7/2} r^{7/2} \sqrt{\frac{2m-r}{m}} sin^{-1}\Big(\frac{\sqrt{r}}{\sqrt{2} \sqrt{m}}\Big) \\&&
+ 72 m^7 sin^{-1}\Big(\frac{\sqrt{r}}{\sqrt{2} \sqrt{m}}\Big)^2 -36 m^6 r sin^{-1}\Big(\frac{\sqrt{r}}{\sqrt{2} \sqrt{m}}\Big)^2\Bigg]
\end{eqnarray}

For all $a \neq 0$, Eq. (\ref{wormholecond2}) is solved numerically to show the required form of $\alpha(r)$ for which the collapse ends up forming a wormhole. For the two signs of $e^{\pm 4 f(r)}$ and different values of mass parameter $m$ the numerical solutions vary slightly. However, they produce a qualitatively similar evolution for different values of $a$. We plot the two evolutions in Fig. \ref{fig2}. \\

To discuss the nature of matter distribution within the collapsing star, we write the nonzero components of the Einstein tensor as
\begin{eqnarray}\nonumber
&& G^0_{\;0} = -\frac{2\zeta\zeta^{\prime}}{r} - \frac{\zeta^2}{r^2},~~ G^1_{\;1} = -\frac{2\zeta\zeta^{\prime}}{r} - \frac{\zeta^2}{r^2} - \frac{2\dot{\zeta}}{r}, \\&&\nonumber
G^1_{\;0} = \frac{2\zeta\dot{\zeta}}{r}, ~~ G^2_{\;2} = G^3_{\;3} = -\frac{\dot{\zeta}+2\zeta\zeta^{\prime}}{r} - \dot{\zeta}^{\prime} - \zeta\zeta^{\prime\prime} - \zeta^{\prime\,2}. \label{FE_gen}
\end{eqnarray}

The stress-energy tensor come into the setup via the field equations $G^{\alpha}_{\;\beta} = -8\pi GT^{\alpha}_{\;\beta}$. Continuity of the energy density can be ensured through the requirement
\begin{eqnarray}
\rho = T^0_{\;0} 
&=& \left\{
\begin{array}{cl}
0, & \textrm{exterior,} \\
\frac{3\alpha \beta^{2}}{\pi G} r^3 [\delta - g T(t)^{n}], & \textrm{interior,}
\end{array} \right.
\end{eqnarray}

Moreover, to ensure a continuity of $\zeta$ as in Eq. (\ref{GeneralFluidpsi}) across the fluid surface
\begin{equation}
r^{3/2} + \frac{1}{(2\beta)^{1/2}}\left(\frac{R}{\alpha}\right)^{1/4} (\delta - gT^n)^{-1/4} = 0. \label{UniformFluidBoundary_gen}
\end{equation}

For a collapsing spherical star that starts shrinking somewhere in negative times and reaches its end state around $t \sim 0$ we approximate Eq. (\ref{UniformFluidBoundary_gen}) by keeping only the terms linear in $t$ and expanding in series. The approximation leads to a simplified continuity equation

\begin{equation}\label{geodesicspecial}
r^{3/2} + \frac{5C_{1}R^{\frac{1}{4}}}{16\beta^{\frac{3}{2}}(\alpha\delta)^{\frac{1}{4}}} \left\lbrace t-\frac{2}{5\beta C_{1}} \left(\frac{\delta n^{2} C_{1}^2}{4} - 5\right)\right\rbrace = 0.
\end{equation}

\begin{figure}
{\centerline{{\scalebox{0.80}{\includegraphics{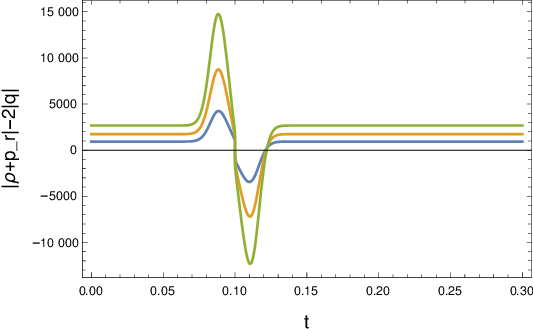}}}}}
{\centerline{{\scalebox{0.80}{\includegraphics{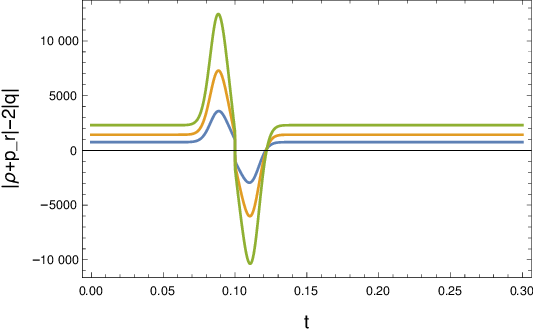}}}}}
\caption{\small \label{fig3} Null Energy Condition for a collapsing fluid sphere in river model. The $\alpha(r)$ profile leading to a bounce and dispersal is chosen. Curves of different color shows the evolution of energy condition for different collapsing shells labeled by different values of $r$.}
\end{figure}

Using the metric Eq. (\ref{gen_PG}), it can be proved that the geodesic equation for a zero energy falling particle in the Schwarzschild exterior region is
\begin{eqnarray}
r^{3/2} + \frac{3\sqrt{R}}{2} (t-t_0) = 0.
\label{PGGeodesicSolution_gen}
\end{eqnarray}

This is easily comparable with Eq. (\ref{geodesicspecial}). If $\alpha(r)$ and the parameters satisfy the following constraints
\begin{equation}
R = \frac{0.002 C_{1}^4}{\alpha\delta \beta^6} ~,~ t_{0} = \frac{2}{5\beta C_{1}} \left(\frac{\delta n^{2} C_{1}^2}{4} - 5\right),
\end{equation}

they match exactly. Therefore, in this parameter range a freely falling particle will hover at the infalling surface of the spherical star. If there is a stress in the surface layer, the internal pressure of the fluid can be balanced and the star is stable. Due to a surface tension the surface can fall faster than a freely falling particle. Around the time the star reaches its minimum accessible volume, irrespective of an onset of it bounce/wormhole throat formation or a forever collapse, the surface tension should become negligible. As a result the system can behave like an idealized pressureless fluid.

\begin{figure}
{\centerline{{\scalebox{0.80}{\includegraphics{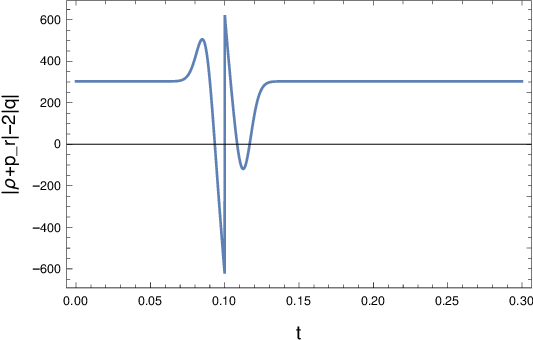}}}}}
{\centerline{{\scalebox{0.80}{\includegraphics{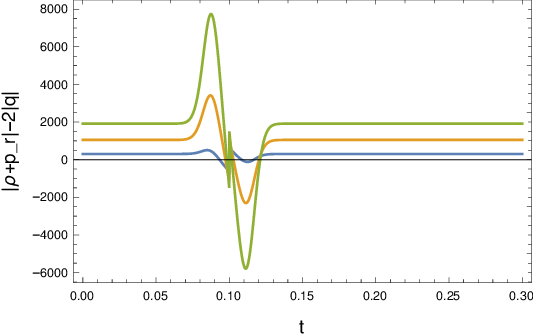}}}}}
\caption{\small \label{fig4} Null Energy Condition for a collapsing fluid in river model. The $\alpha(r)$ profile leading to a Wormhole outcome is chosen. Curves of different color shows the evolution of energy condition for different collapsing shells labeled by different values of $r$.}
\end{figure}

It is expected under normal circumstances that the energy-momentum tensor components will satisfy certain energy conditions during the gravitational collapse and maintain a notion of `locally positive energy density'. We focus in particular, on the Null Energy Condition (NEC) which has a more generic root in the Null Convergence Condition \cite{ncc}. It has an algebraic form written as  
\begin{equation}\label{ec1}
NEC \rightarrow \mid \rho + p_{r} \mid- 2\, \mid q\mid \geq 0.
\end{equation}

We plot the NEC as a function of time, for different values of $r$, i.e., distance from the center of the sphere. $r = 1$ is taken as the boundary of the star. Fig. \ref{fig3} shows the evolution of NEC with time as the collapsing fluid bounces and disperses away all clustered matter. There is a clear violation of NEC at the critical point as the bounce starts. For this plot the analytical solution of $\alpha(r)$ as in Eq. (\ref{alphaform}) is used. In Fig. \ref{fig4}, we plot the NEC profile as the collapsing sphere develops a wormhole throat, using numerical solution of Eq. (\ref{wormholecond2}). It is seen that not only the NEC is violated, there is a curious onset of periodicity during the wormhole throat formation. \\

In summary, this letter provides an alternative picture of gravitational collapse using the unconventional set of coordinates, namely, the generalized PGL metric. We name it a \textit{river model}, owing to the illustration of space flowing with Newtonian escape velocity, much like a river, through flat background. While the static PGL metric is known for almost a century, we construct a new time-evolving PGL geometry describing gravitational collapse of a spherical stellar body. We prove that a spherically symmetric metric should satisfy a set of transformation equations in order to have a PGL form and find a unique, exact solution to this set. This solution can describe three possible non-singular outcomes, depending on initial conditions. The first two are either a collapse for infinite time without any zero proper volume, or a bounce and eventual dispersal of all the clustered matter. It is curious to find that there is also a clear third alternative, where the collapsing body forms a wormhole throat at some finite value of the radial coordinate. This is proved by exploring the interior geometry on an Euclidean metric embedded in a three-dimensional space. As an example, we discuss the requirement for the formation of a Simpson-Visser wormhole geometry, which falls within a wider class of static geometries behaving as Black Hole mimickers \cite{mimic}. Therefore, the consequence of a gravitational collapse in the river-frame is quite intriguing; it implies that just a simple requirement to allow a consistent river metric/PGL form ensures a resolution of the singularity problem even within the realms of classical general relativity. The original static metric and the dynamic counterpart, both are motivated from the analogy that a sonic barrier in fluid dynamics can behave like a general relativistic event horizon \cite{unruh, unruh1}. The static geometry has remarkably led to laboratory simulations of sonic black holes in some very realistic states of matter, such as Bose-Einstein Condensates \cite{lahav}. It is our optimism that the time-evolving twin will also be of immense use in laboratory simulations of gravitational collapse, which is perhaps the only phenomena through which a classical stellar body can evolve naturally towards a quantum scale. Furthermore, using the exact metric prescribed in this letter, simulating the sonic analogue of a wormhole (\textit{not-so-dumb hole}) for the very first time, also seems plausible for a reasonably well-defined fluid flow.

\end{document}